\documentclass[prL,twocolumn,showpacs,amsmath,amssymb]{revtex4}
\usepackage{graphicx}
\usepackage{epstopdf}

 \def \In {$^{115}$In }
 \def\Ce {CeCoIn$_5 $ }

\begin{document}

\title{Field Dependence of the Ground State in the Exotic Superconductor CeCoIn$_5$:\\
a Nuclear Magnetic Resonance Investigation}
\author{ G. Koutroulakis$^{1}$, V. F. Mitrovi{\'c}$^{1}$, M. Horvati{\'c}$^{2}$, C. Berthier$^{2}$,
G. Lapertot$^{3}$, and J. Flouquet$^{3}$}
\address{$^{1}$Department of Physics, Brown University, Providence, RI 02912, U.S.A. \\
$^{2}$Grenoble High Magnetic Field Laboratory, CNRS, B.P. 166, 38042 Grenoble
Cedex 9, France\\
 $^{3}$D{\'e}partement de Recherche Fondamentale sur la Mati{\`e}re Condens{\'e}e, SPSMS, CEA Grenoble, 38054 Grenoble Cedex 9, France}

\date{Version \today}

\begin{abstract}

We report \In nuclear magnetic resonance (NMR) measurements in \Ce at low
temperature ($T \approx 70$~mK) as a function of magnetic field ($H_0$) from
 \mbox{2 T} to \mbox{13.5 T} applied perpendicular to the $\hat c$-axis. NMR line shift
reveals that below 10~T the spin susceptibility increases as $\sqrt{H_0}$. We
associate this with an increase of the density of states due to the Zeeman and
Doppler-shifted quasiparticles extended outside the vortex cores in a $d$-wave
superconductor. Above 10 T a new superconducting state is stabilized, possibly
the modulated phase predicted by Fulde, Ferrell, Larkin and Ovchinnikov (FFLO).
This phase is clearly identified by a strong and linear increase of the NMR
shift with the field, before a jump at the first order transition to the normal
state.
\end{abstract}

\maketitle


Applying a magnetic field to a superconductor is a powerful way of revealing
the complexity of this macroscopic quantum state of matter. Even though the
effects of the field in conventional type-II superconductors are well
established, in a wide range of systems (such as the high-$T_c$, heavy
fermions, and organic superconductors) the consequences are far from being
understood. One possibility is that the magnetic field may affect the
competition between superconductivity and antiferromagnetism (or other
competing orders), as suggested by several experiments and theories in
high-$T_c$ superconductors \cite{Lake05, Sachdev02}. Particularly in the case
of a $d$-wave superconductor, study of the low energy excitations in the vortex
state can provide useful information about the nature of the underlying ground
states. The local density of states (DOS) of these excitations can be
effectively probed by measuring the magnetic field dependence of the NMR shift
as presented here. Single crystals of heavy-fermion superconductor \Ce are an
ideal material for such a study owing to their high purity, most likely
$d$-wave gap symmetry \cite{Kohori, Kawasaki03}, and entirely accessible
magnetic field - temperature ($T$) phase diagram.

Furthermore, an applied magnetic field ($H_0$) may induce a novel
superconducting (SC) state in which the momentum of the Cooper pairs is not
equal to zero, but becomes finite and proportional to $H_0$, as predicted by
Fulde, Ferrell, Larkin, and Ovchinnikov (FFLO) \cite{FFLOdis}. In this state
the SC order parameter oscillates in real space. The FFLO phase is expected to
occur in the vicinity of $H_{c2}$ when Pauli pair breaking dominates over
orbital effects. The search for this exotic SC phase has attracted much of
current interest \cite{Bianchi03, Radovan03, Uji06, Lortz07}, and there is experimental
evidence that it is realized in CeCoIn$_5$ \cite{Bianchi03, Radovan03, Martin05,
mitrovic06}. Moreover, CeCoIn$_5$ is   close to an antiferromagnetic (AF) instability,
offering thus a unique possibility to study the competition between FFLO and AF
instabilities. It was recently shown that a substitution of \mbox{10 \%} of Cd
for In could generate AF droplets \cite{Urbano07}. Even in pure CeCoIn$_5$, NMR
has revealed the existence of some form of magnetism within the possible FFLO
phase \cite{Young07}.

Here we report detailed \In NMR measurements as a function of $H_0$ from
\mbox{2 T} to \mbox{13.5 T} at \mbox{$T = 73$ mK}. We establish that the shape
of the NMR spectra is determined by the inhomogeneous field distribution of a
vortex lattice (VL) for \mbox{$H_0 \lesssim 10$ T}.  For  fields above \mbox{10 T}, where a possible FFLO
state (henceforth referred to as ``modulated'' or mSC state) is stabilized, the
NMR spectra broaden well beyond what is expected on the basis of the VL
distribution.
Magnetic field dependence of the shift reveals explicit effects of the Zeeman
energy on quasiparticles in this Pauli-limited superconductor. Below 10 T, the
spin susceptibility increases as $\sqrt{H_0}$. This increase is related to the
excess DOS due to the Zeeman and Doppler-shifted nodal quasiparticles,
confirming the presence of gap nodes in the SC state of CeCoIn$_5$. Above 10 T,
the spin susceptibility is strongly enhanced and increases linearly with $H_0$,
as expected in the FFLO state \cite{Vorontsov06}. If instead this enhancement
is ascribed to field induced magnetism, our data place a lower-bound on the
spatial extent of the magnetic regions.

We have used high quality single crystals of \Ce grown by a flux method
\mbox{\cite{CedaDic}}.  
Here, the
 spectra of the axially symmetric In(1) site in $H_0 \, ||\, [100]$ (aligned to better than 2$^\circ$) are
reported. They were recorded using a custom built NMR spectrometer and
obtained, at each given value of $H_0$, from the sum of spin-echo Fourier
transforms recorded at constant frequency intervals. The magnetic shift was
determined by the diagonalization of the full nuclear spin Hamiltonian and by
subtracting the orbital contribution of $\simeq$~0.13~\%. The RF coil was
mounted into the mixing chamber of a $^3$He/$^4$He dilution refrigerator while
a variable tuning delay line and matching capacitor were mounted outside the
NMR probe to allow for a wide frequency/field coverage. The $^{63}$Cu NMR on
copper nuclei from the RF coil was used to determine the exact value of $H_0$.
The sample was both zero-field and field-cooled and no visible influence of the
sample's cooling history on the NMR spectra was detected. In order to avoid
heating of the sample by RF pulses we used very weak RF excitation power
\cite{MitrovicConfP} and repetition times of the order of ten seconds.

\begin{figure}[t]
 \begin{minipage}{0.98\hsize}
\centerline{\includegraphics[width=8.0cm]{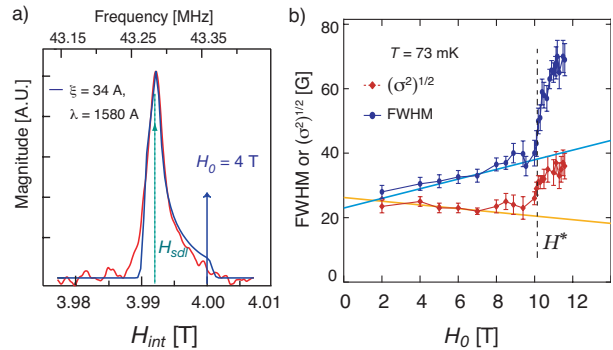}} 
\begin{minipage}{0.98\hsize}
  \vspace*{-0.1cm}
\caption[]{\label{Fig1}(Color online)  \small {\bf a)}   In(1), $\langle - 1/2 \leftrightarrow - 3/2 \rangle$
transition, spectra (red) at
\mbox{$T = 73$ mK} and \mbox{$H_0 = 4$ T} compared to calculated VL lineshape
(blue), as described in the text, as a function of the internal magnetic field
($H_{int}$). The arrows indicate $H_{int}$ corresponding to the saddle point
($H_{sdl}$) and $H_0$ fields, respectively. {\bf b)}  
FWHM and $\sqrt{\sigma^2}$  of the
measured In(1) spectra (filled symbols) as a function of $H_0$. Solid lines are
the calculated values obtained as described in the text. The dashed line
denotes the field $H^*$.}
  \vspace*{-0.4cm}
 \end{minipage}
 \end{minipage}
\end{figure}

In the Abrikosov SC state (henceforth referred to as the ``uniform'' uSC
state), at low $T$, vortices tend to form a solid lattice \cite{Abrikosov57}
resulting in a spatial distribution of magnetic fields. This distribution is
reflected in the NMR lineshape, as shown in \mbox{Fig. \ref{Fig1}a)}, and can
be calculated by solving the Ginzburg-Landau (GL) equations \cite{Abrikosov57,
Brandt97}. We calculated the spectra employing Brandt's method \cite{Brandt97},
valid at any value of  $H_0$ between the lower and upper critical fields.
The input parameters for the calculation are the $H_0$, coherence length ($\xi$), 
penetration
depth ($\lambda$), and
VL geometry (the angle $\alpha$). For a given value of $H_0$, a $\chi^2$
minimization of the difference between the calculated and measured spectra,
normalized to their corresponding areas, was performed with $\xi$, $\lambda$,
and $\alpha$ as variational parameters. As the result is not very sensitive to
the variation of $\alpha$ parameter, its value was set to $\alpha = 90 ^\circ$,
{\it i.e.} a square vortex lattice. At \mbox{$H_0 = 4$ T} the minimum $\chi^2$
was achieved for  \mbox{$\xi= 34 \pm 10$ \AA}\, and  
  \mbox{$\lambda = 1580 \pm 120$ \AA}, values consistent with earlier
reports \cite{Ormeno02}. The spectra were then simulated for all field values
$2\, {\rm T} < H_0 < 11.6\, {\rm T}$ using this {\it single} set of parameters.
 In order to make a
 quantitative comparison, the full width at half maximum
(FWHM) and the square root of the second moment ($\sqrt{\sigma^2}$) of the
simulated and measured spectra are plotted in \mbox{Fig. \ref{Fig1}b)} as a
function of $H_0$. The agreement between the simulation and the experiment in
the uSC state for \mbox{$H_0 \lesssim 10$ T} is excellent. To our knowledge
this is the first time that the comparison is performed over such a wide field
range. Within the sensitivity of our measurements, these results indicate that
for $H_0 || [100]$, ranging from \mbox{2 T} to \mbox{10 T}, the GL model 
provides an adequate
description of the vortex state even for this Pauli-limited superconductor,
contrary to the results of neutron scattering study for $H_0$ perpendicular to
the planes \cite{Bianchi07}. This discrepancy might stem, besides $H_0$  orientation,  from the fact that the probability distribution, {\it i.e.} NMR spectrum, is  insensitive to the  Pauli effects    which are enhanced near the vortex cores \cite{Ichioka07}. 

For fields above 10 T, the measured spectra broaden beyond what is expected
from the calculated VL distribution in the GL model. We have found that it is
impossible to account for the high field linewidth broadening regardless of the
$\xi$ and $\lambda$ values used. An appreciable discrepancy begins at
\mbox{$H_0 = H^*$}  that  corresponds to the transition field from the uSC to the mSC state
\cite{Bianchi03,MitrovicConfP}. Therefore, we demonstrate that at low $T$ the
transition to the mSC state is identified by the onset of the additional
asymmetric NMR linewidth broadening. This implies the presence of an additional
modulation of the internal field  in the mSC state. The lineshape contains a
{\it single} peak, which is consistent with an incommensurate modulation of the
internal field along {\it two} spatial directions \cite{Berthier76}, as
expected for an FFLO state in a $d$-wave SC \cite{Maki02, Wang06}.

\begin{figure}[t]
 \begin{minipage}{0.98\hsize}
\centerline{\includegraphics[width=8.0cm]{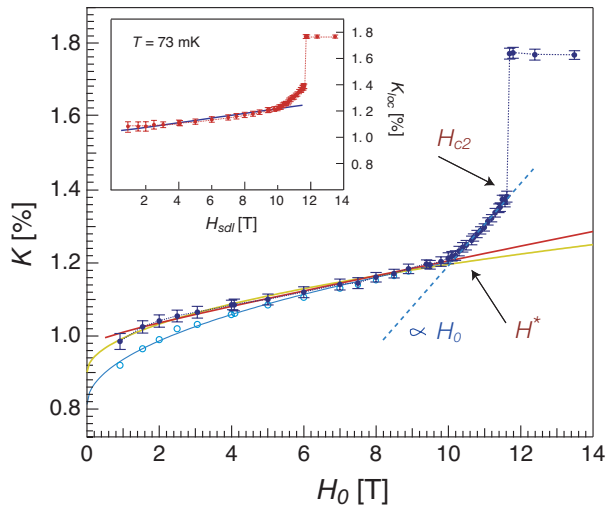}} 
\begin{minipage}{0.98\hsize}
 \vspace*{-0.3cm}
\caption[]{\label{Fig3} (Color online)  \small In(1) magnetic shift (open circles) as a function
of $H_0$ at \mbox{$T = 73$ mK}. The solid circles are the shift corrected for
diamagnetic contribution, {\it i.e.} shift with respect to the average magnetic
field in the VL unit cell.   The dotted lines
are guide to the eye. The darker (red) solid line is the calculated shift, as
described in the text.
 The lighter
solid lines are a fit of the data, for \mbox{$H_0 < 10$ T}, to \mbox{Eq.
\ref{sFit}}. The dashed line is a linear fit to the data for \mbox{$H_0 > 10$
T}. Inset: Magnetic field dependence of the local In(1) shift $K_{loc}$ at the
saddle point field. The solid line is calculated shift, $K= 1.04 \%+ 77.1 \cdot
(\epsilon + E_Z + E_{Dsdl}) \, \% {\rm eV}^{-1}$. }
 \end{minipage}
 \end{minipage}
 \vspace*{-0.4cm}
\end{figure}

We now proceed to the discussion of the main result, the $H_0$ dependence of
the low $T$ shift data illustrated in \mbox{Fig. \ref{Fig3}}. Two phase
transitions are clearly discerned. The first order phase transition from the
normal to the mSC state at $H_0 = H_{c2}$ is characterized by the sizable
discontinuity of the shift. The continuous second order phase transition from
the uSC to mSC state occurs at \mbox{$H_0 = H^*$}. Presented data provide the
first clear NMR signature of this phase transition
\cite{MitrovicConfP,Young07}. In both SC states, the shift increases with
increasing $H_0$. However, above $H^*$ the field dependence of the shift is
significantly enhanced. In the following, we will first address variation of
the shift in the uSC.

In a SC with the $d$-wave gap symmetry in the excitation spectrum, at low $T$
the quasiparticle excitations are restricted to nodal regions \cite{Volovik}.
These low energy quasiparticles can be probed with NMR shift measurements.
Specifically, the shift is proportional to the DOS ($N(E )$) averaged
over energy $E$ in a range of the order of $k_B T$ around the Fermi
energy, $E_F$ \cite{AbragamBook}. Thus, at sufficiently low $T$ the shift is
proportional to the DOS at the Fermi level, {\it i.e} $K \sim \langle
N(E )\rangle \equiv N_F$. This implies that in the SC state only the DOS
in the regions around the nodes of the gap contributes to $K$. Near the nodes
the DOS depends linearly on quasiparticle excitation energy, for  energies
less than half of the gap magnitude. Therefore, the low $T$ shift is given by,
\begin{equation}
 \label{eqKen}
K \sim \langle N_F \rangle \propto \langle | \epsilon + E_Z + E_D| \rangle,
\end{equation}
 where $\langle ... \rangle$ denote the average over four nodes in  the $k$-space,
 and $\epsilon \approx k_BT$, \mbox{$E_Z=   \mu_{eff} H_0$}, and \mbox{$E_D={\bf v}_{F} \cdot {\bf p}_s$} are
 the thermal, Zeeman, and Doppler
energy terms, respectively. The Doppler term originates from the shift of the
excitation energies of the nodal quasiparticles, with Fermi velocity ${\bf
v}_{F}$, moving in the superflow field with momentum ${\bf p}_s$. Both $E_Z$
and $E_D$ depend on $H_0$. The $H_0$ dependence of the local ${\bf p}_s$, and
thus $E_D$ at any point in the real space of the VL, can be calculated
exploiting Brandt's algorithm \cite{Brandt97}. The same algorithm is then used
to directly calculate the $H_0$ dependence of the shift averaged over the real
space of a VL unit cell from \mbox{$K= K_0 + K_e (\left \langle|\epsilon + E_Z
+ E_{D}|\right\rangle)$}. The following input parameters were used: ${\bf
v}_{F} \hbar = 49.1 \, {\rm meV \AA}$ \cite{Miclea06}, \mbox{$E_Z=  0.5 \mu_B H_0$}, $K_0 = 0.97\, \%$, and
\mbox{$K_e = 85.24 \, {\rm \%eV}^{-1}$}, determined so that $K$ at $H_{c2}^{\rm
orb}$, the orbital \mbox{$H_{c2} = 38.6$ T}, equals the normal state shift.
Excellent agreement with the data for $H_0 < 10$ T is obtained with this
simple model, as illustrated in \mbox{Fig. \ref{Fig3}}. 

We point out that
the shift   has been determined from the frequency corresponding to the first
moment, that is the average frequency, of the spectra and thus reflects the
real space average of $K$ over a VL unit cell. In \mbox{Fig. \ref{Fig3}} we
have thus presented both the shift data with respect to the applied field
$H_0$, as well as the same data calculated with respect 
to the field averaged over the VL unit cell
 as  obtained from the Brandt's algorithm. 
 In our calculation of the average $K$ in the uSC state, we neglected the Pauli paramagnetic effects and their variation across the VL. This is justified by the   fact  that these   effects are insignificant outside  and are enhanced only near the vortex cores. 
 Thus, mainly due to the large volume
contribution from the outside of the cores,
the average DOS is not notably affected by
the Pauli paramagnetism  \cite{Ichioka07}.  
 The field dependence of
the data is fully explained by the increase in the dominant Zeeman and average
Doppler energy of the quasiparticles. Note that this behavior is in sharp
contrast with that in a conventional SC without nodes, where the Doppler term
has a negligible effect. Thus, our results indicate that \Ce is a SC with nodes
in the gap, which is most-likely of $d$-wave ($d_{x^2-y^2}$) symmetry. 

The observed field dependence is also consistent
 with Volovik's
prediction \cite{Volovik} for a $d$-wave SC that the average DOS $\propto \sqrt
{H_0 /H_{c2}}$. 
As shown in \mbox{Fig. \ref{Fig3}}, \mbox{$H_0 < 10$ T} data is well fitted to
\begin{equation}
\label{sFit}
K = K_0 + \beta K_n \sqrt{H_0/H_{c2}^{\rm orb}},
\end{equation}
where $H_{c2}^{\rm orb}$  is fixed to 38.6 T and
\mbox{$K_n = 1.77 \, \%$} is the normal state shift. The fit parameters are
$K_0 = 0.90 \pm 0.01 \, \%$ and $\beta = 0.33 \pm 0.02$. In the
limit $H_0 \rightarrow 0$, $K$ remains finite, at nearly half of the normal
state shift. The finite shift, that is the field independent constant
contribution $K_0$, may be attributed to the multi-band nature of SC in
CeCoIn$_5$. The contribution $K_0$ would then  originate from   {\it normal} quasiparticles in the small gap band
\cite{Seyfarth}.

In order to verify the significance of the contribution of $E_D$ to the
quasiparticle energy, we have also extracted the \textit{local} shift from the
peak of the spectra which corresponds to the nuclei positioned at the saddle
point of the field distribution, that is at the point in real space midway
between two vortices. At this point where $H = H_{sdl}$, the local ${\bf p}_s$,
and thus $E_D$, increases very slightly with increasing $H_0$ (for $H_0 \ll
H_{c2}^{orb}$) since it is not influenced by geometry effects such as the
change in the vortex number with varying $H_0$. Thus, the local shift at the
sadle point should exhibit weaker field dependence than the shift averaged over
the entire VL. This is indeed the case as shown in the inset to \mbox{Fig.
\ref{Fig3}}, where the local shift at $H_{sdl}$ is displayed. This shift corresponds to that  
of the peak of the spectrum calculated with respect to the $H_{sdl}$, obtained
from Brandt's algorithm. Here the observed field dependence arises solely from
the increase of the quasiparticles Zeeman energy.

We next consider the field dependence of the shift in \mbox{$10 T < H_0 < H_{c2}$}. As
apparent in \mbox{Fig. \ref{Fig3}}, in this regime $K$ increases linearly with 
  $H_0$. Contrary to \cite{Radovan03}, no evidence of discontinuous jump in $K$, indicating  the  transitions between different Landau levels within the FFLO state,  is observed. 
Our data can be well described by
\begin{equation}
K = K^h_0 + \beta_h K_n \left( H_0/ H_{c2} \right),
\end{equation}
where $H_{c2}$ is fixed to 11.7 T, $K_n = 1.77 \, \%$ is the normal state
shift, \mbox{$\beta_h = 0.77 \pm 0.02$}, and $K^h_0 = 0.03 \pm 0.01$.
 The latter parameter shows that as $H_0 \rightarrow 0$, $K$ extrapolates to zero.
The rate of increase of the shift $(\Delta K/\Delta H_0)$ is five times higher than that in the uSC
for $H_0 < H^*$. After careful consideration of all the possibilities regarding
the values of $K^h_0$ and the Doppler term, we conclude that this large
increase can be ascribed to an enhancement of $E_Z$ and its dominance over
$E_D$ in the entire VL unit cell.
For several reasons the fulfillment of  $E_Z > E_D$ condition  is very likely
  in the FFLO state. First, the internal
field, due to paramagnetic effects, can be large enough in the FFLO state to
exceed the $E_D$ term everywhere in the VL unit cell. Second, it is possible
that  $E_D$ is suppressed by either the opening of a subdominant  gap, or a change in
the gap structure so that ${\bf v_F}$ and ${\bf p_S}$ become nearly orthogonal.
The gap opening scenario is very unlikely since $K$ increases in the FFLO state
contrary to expectations in the presence of a gap \cite{Wang06, Maki02}. On the
other hand, changes in the gap structure are  expected in the FFLO state.
The observed $\Delta K/ \Delta H_0$ could reflect the fact that the DOS is no
longer simply proportional to $E$. That is, the DOS dispersion relation changes
and/or an extra structure, such as bound states, is induced by $H_0$. This is
consistent with the FFLO scenario \cite{Wang06} but further  
calculations are required for a   quantitative comparison.

Nonetheless, a magnetic origin of the mSC phase or  coexistence of some magnetic order with the FFLO
  state cannot  be excluded
\cite{Young07, mitrovic06}. Given  the  {\it a priori} antagonistic relationship between SC and magnetism,
 it is likely that magnetism   appears  in the spatial regions where   SC is suppressed as is the case in the vortex cores \cite{Lake05, Young07}. However, it  was shown that the existence of the local moment magnetism requires pair coherence \cite{mitrovic06}.
The absence of magnetism in the normal state could be attributed to
  the Kondo screening, acting on a short length scale,  of Ce local moments.
In the SC state the Kondo screening becomes ineffective, since quasiparticles
have very small momentum inhibiting short range screening,   giving  rise to
 magnetism. Besides,  a  magnetic phase could be stabilized only in high
fields when there is a sufficient overlap between vortex cores, so that the
correlations between magnetic regions can be established.  In  field of
10~T, the distance between neighboring vortices is \mbox{$\approx  140$ \AA}.
Thus,  the long range magnetic order could be established in fields above 10 T
if  the radius of the magnetic cores is larger than \mbox{$2\xi \approx 70$
\AA}. In this case,  the field enhancement of the shift would be a consequence
of the canting of local Ce moments. Varying $H_0$ from 10.2~T to 11.7~T
increases the shift from 1.22~\% to 1.38~\%. With a hyperfine coupling between
In(1) nucleus and its 4 Ce neighboring atoms of $A\approx 1.2$~T/$\mu_B$
\cite{Curro01}, the increase of the average local magnetic moment is $\Delta
\mu_{eff} = \Delta(K H_0)/4A \approx 0.0075 \, \mu_B/{\rm Ce}$. This
corresponds to 7.5~\% of a typical
 normal state local Ce moment ($\simeq 0.1\, \mu_B$), implying that for $H^* < H_0 <
H_{c2}$, the increase of $H_0$ induces a canting of  \mbox{$\simeq 7.5 \, \%$} of the normal state  local Ce moment.

In conclusion, our essentially zero-$T$ limit data provide the clearest NMR
evidence of two phase transitions in the vicinity of $H_{\rm c2}$ thus far. In
the uSC phase, the NMR spectra can be nicely fitted to the calculated magnetic
field distribution arising from the vortex supercurrents with SC coherence
length \mbox{($\xi \approx 34$ \AA)}\, and penetration depth \mbox{($\lambda
\approx 1580$ \AA)} as fitting parameters. 
To explain the
observed increase of the spin susceptibility with $H_0$, we conclude that the
dominant low energy excitations are the Zeeman and Doppler-shifted
quasiparticles extending outside the vortex cores, implying that \Ce is of
$d$-wave gap symmetry. For the high-field  (above 10.2~T)  low-$T$ phase we find
that it cannot correspond to a simple VL rearrangement. It is consistent with
an FFLO state with 2D incommensurate modulation of the quasiparticle density in which the spin susceptibility strongly increases 
as $H_0$.
If the magnetism appears there in the spatially localized fashion, we show that
magnetic cores extend on a length scale  larger than $2\xi$.

We are very grateful to V. Mineev,   M. Eschrig, and  S. Kr{\"a}mer for enlightening discussions.
supported by the funds from NSF (DMR-0710551), ANR grant 06-BLAN-0111, and the
GHMFL, under European Community contract RITA-CT-2003-505474. V. F. M.
acknowledges support by the A. P. Sloan Foundation.

\bibliographystyle{unsrt}

\vspace{0.0cm}
\end{document}